\documentclass[preprint,showpacs,showkeys,preprintnumbers,amsmath,amssymb]{revtex4}

\usepackage{graphicx}
\usepackage{dcolumn}
\usepackage{bm}

\begin{document}


\title{Critical dynamics and global persistence exponent on Taiwan financial market}

\author{I-Chun Chen}
\email{ichun.nancy@msa.hinet.net}
\affiliation{Physics Department, National Chung Hsing University, Taichung, Taiwan R.O.C.}
\affiliation{Nan Kai Institute of Technology, Nantou, Taiwan R.O.C.}

\author{Hsen-Che Tseng}
\author{Ping-Cheng Li}
\author{Hung-Jung Chen}
\affiliation{Physics Department, National Chung Hsing University, Taichung, Taiwan R.O.C.}


\begin{abstract}
We investigated the critical dynamics on the daily Taiwan stock
exchange index (TSE) from 1971 to 2005, and the 5-min intraday
data from 1996 to 2005. A global persistence exponent $\theta_{p}$ was
defined for non-equilibrium critical phenomena \cite{Janssen,Majumdar}, and
describing dynamic behavior in an economic index \cite{Zheng}.

In recent numerical analysis studies of literatures,
it is illustrated that the persistence probability
has a universal scaling form $P(t) \sim t^{-\theta_{p}}$ \cite{Zheng1}. In this work,
we analyzed persistence properties of universal scaling
behavior on Taiwan financial market, and also calculated
the global persistence exponent $\theta_{p}$. We found our analytical
results in good agreement with the same universality.

\end{abstract}

\pacs{47.27.eb}
\keywords{Persistence probability; Hurst exponent; Taiwan stock exchange index.}
\maketitle

\section{\label{sec:level1}Introduction}

Problems in economy and finance have attracted the interest
of statistical physicists all over the world. Using the
tools developed for statistical physics, like phase
transitions, critical exponents, mean field approximations,
renormalization group \cite{Didier}, persistence
probability \cite{Majumdar1,Li,Constantin}.

In recent years the detrended fluctuation analysis (DFA)
method \cite{Carbonea,Ausloos,Vandewalle,Matia,Oswiecimka,Schmitt}
has become a widely used technique for the
determination of (mono-) fractal scaling properties and
the detection of long-range correlations in noisy,
nonstationary time series \cite{Vandewalle}. In many of non-equilibrium
systems, the persistence has been found to decay as a
power-law at time series, $P(t) \sim t^{-\theta_{p}}$. Hurst exponent and
persistence exponent in these financial time series are
investigated in numerical and analytical \cite{Constantin}.

We calculated the experimental data with the daily Taiwan stock exchange
index (TSE) from 1971 to 2005, and the 5-min intraday data
from 1996 to 2005. In this work, we analyzed persistence
properties of universal scaling behavior on Taiwan financial market.

\section{\label{sec:level2}Method}

We consider a set of data recorded the daily Taiwan
stock exchange index (TSE) from 1971 to 2005.
Let $Y_{i}(t)$ be the stock index at discrete times $i$,
$i = 1, 2,...,t_{n}-1$. The final transaction
time is denoted by $t_{n}$. Then, the log-return price
is defined as
\begin{eqnarray}
r_{i}(t)=\ln{Y_{i}(t+\Delta t)-\ln{Y_{i}(t)}},
\label{1}
\end{eqnarray}
where $\Delta t$ is time interval. In this paper,
we analyzed daily data; $\Delta t=1$ day.

\begin{figure}
\includegraphics{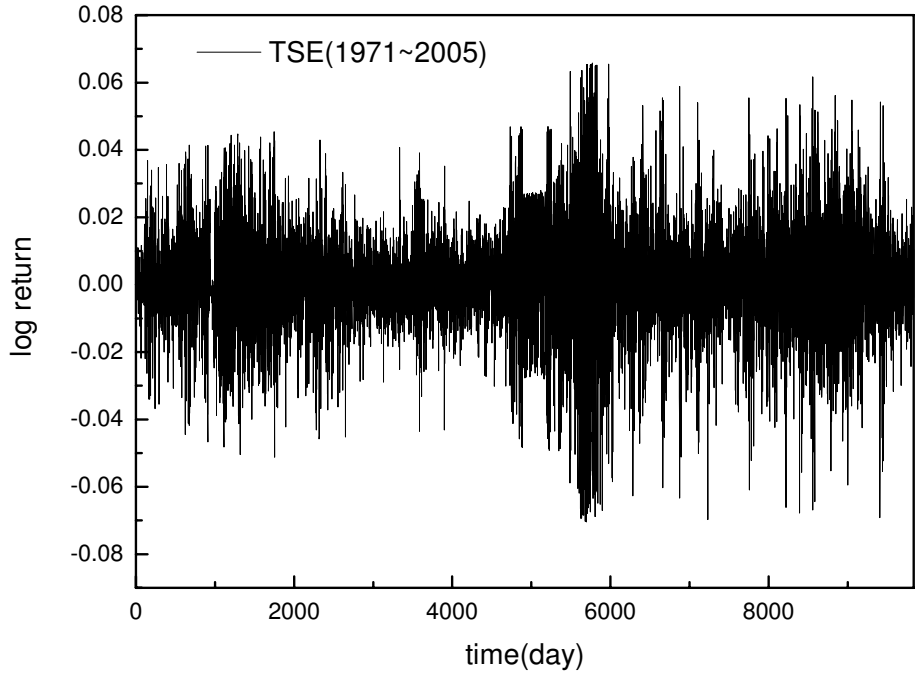}
\caption{\label{fig:1} Plot of the log-return price
$r_i(t)$ vs. $t$ on the daily Taiwan stock exchange
index (TSE) from 1971 to 2005.}
\end{figure}

Let us denote the value of the TSE at a certain time $t'$ as $y(t')$.
$P_{+}(t)$ is the probability that the value of the TSE has never gone down to the
value $y(t')$ in time $t$, $y(t'+N\Delta t) > y(t')$ for $N=1,2,...,t$. i.e.,
$P_{-}(t)$ is the probability that the value of the TSE has never
gone up to the value $y(t')$ in time $t$, $y(t'+N\Delta t) < y(t')$ for $N=1,2,...,t$.
The persistence probability $P(t)$ is $[P_{+}(t)+ P_{-}(t)]/2$.

The persistence probability has a power-law behavior
\begin{eqnarray}
P(t) \sim t^{-\theta_{p}}.
\label{2}
\end{eqnarray}

\begin{figure}
\includegraphics{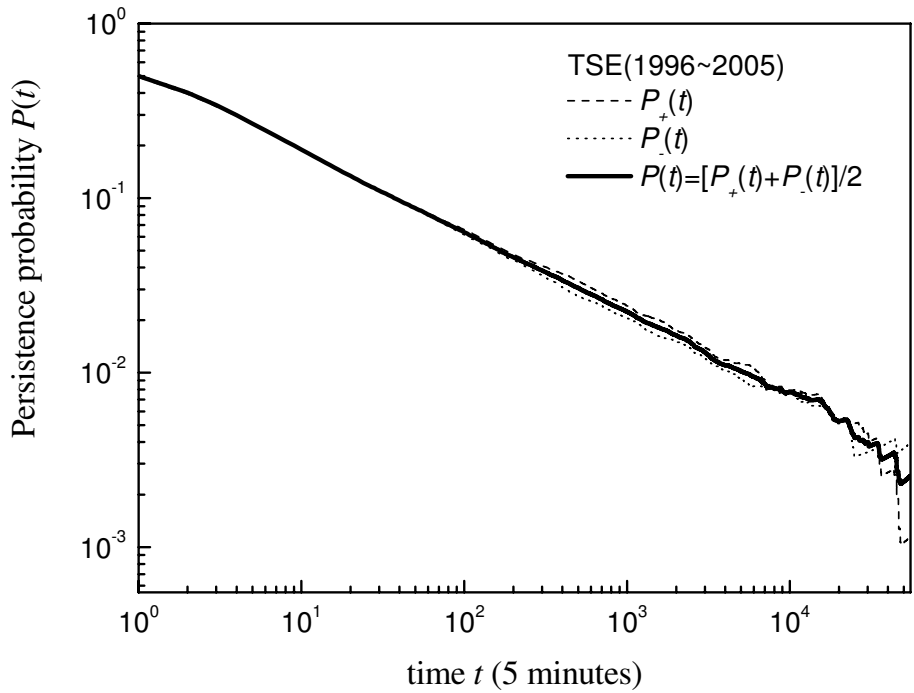}
\caption{\label{fig:2a} Persistence probability with 5 minutes
data of the daily Taiwan stock price index (TSE) from 1996 to 2005.}
\end{figure}

\begin{figure}
\includegraphics{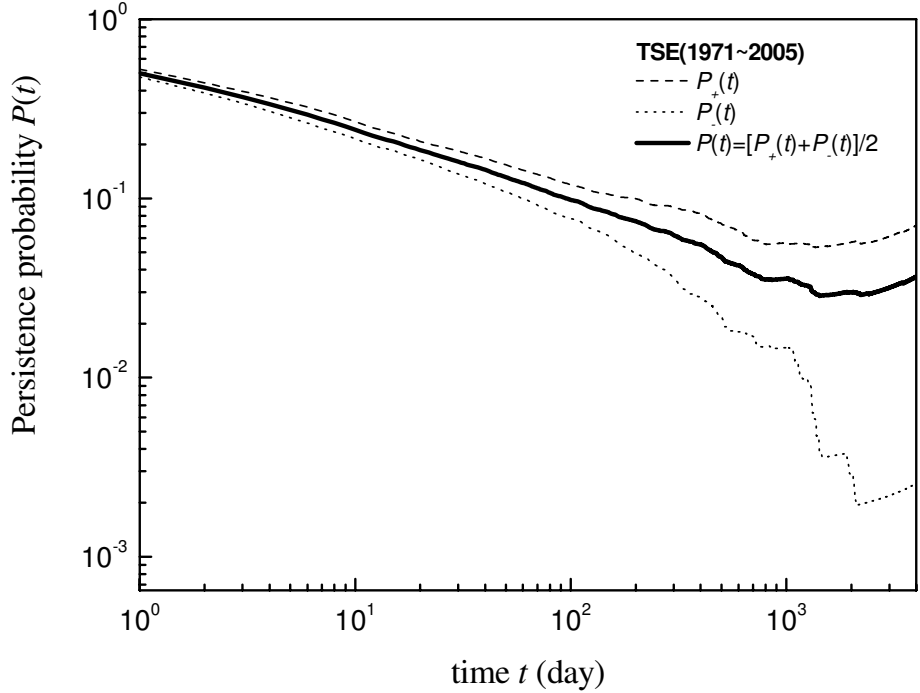}
\caption{\label{fig:2b} Persistence probability with 5 minutes
data of the daily Taiwan stock price index (TSE) from 1996 to 2005.}
\end{figure}

For the cumulative time series of the log-return price variables is defined as
\begin{eqnarray}
X(i)=\sum_{k=1}^{i} (r_{k}-{\bar r} ),
\label{3}
\end{eqnarray}
where ${\bar r}$ is the average value of log-return price.
$X(i)$ is divided into $Ns$ disjoint segments of length $s$.
$p_{\nu}(i)=a_i+b_it$ , $a_i$,  $b_i$ is constant. Since the
length $N$ of the series is often not a multiple of the considered
\begin{eqnarray}
X_{s}(i)=X(i)-p_{\nu}(i).
\label{4}
\end{eqnarray}
The generalized $q$~th-order price-price correlation function is defined as
\begin{eqnarray}
G_q(t)={\langle |Y(t_0+t)-Y(t_0)|^q \rangle}^{1/q},
\label{5}
\end{eqnarray}
where $Y(t)$ is the stock price and the average is over all the initial times $t_0$.
$G_q(t)$ has a power-law behavior
\begin{eqnarray}
G_q(t) \sim t^{H_{q}}.
\label{6}
\end{eqnarray}
where $H_q$ is called the generalized Hurst exponent.

\begin{figure}
\includegraphics{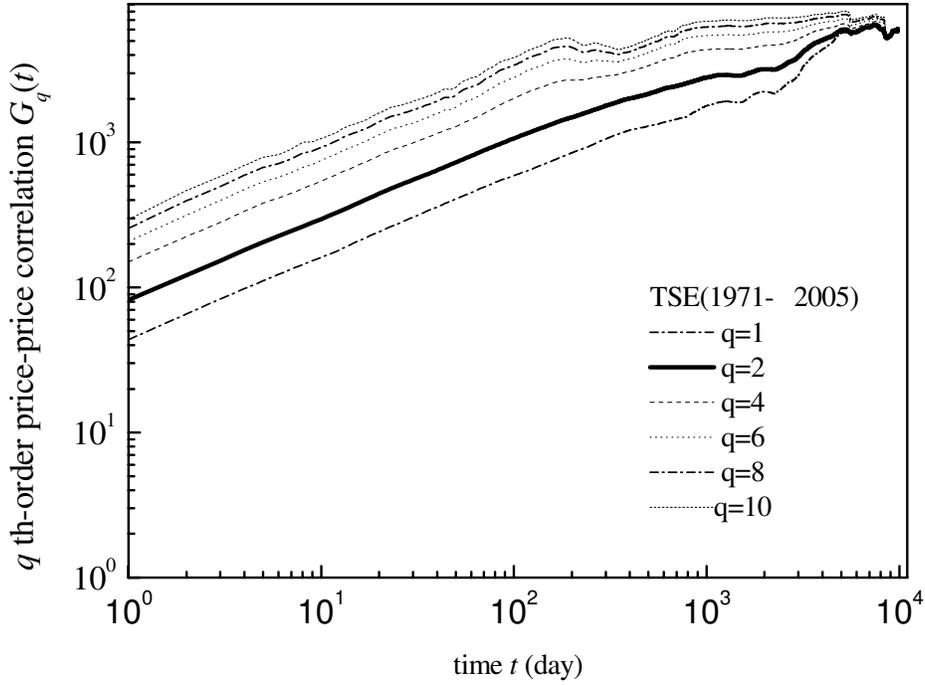}
\caption{\label{fig:3a} Log-log plot of the generalized
price-price correlation function $G_q(t)$ vs. $t$ corresponding
to the daily Taiwan stock exchange index (TSE) from
1971 to 2005. }
\end{figure}

\begin{figure}
\includegraphics{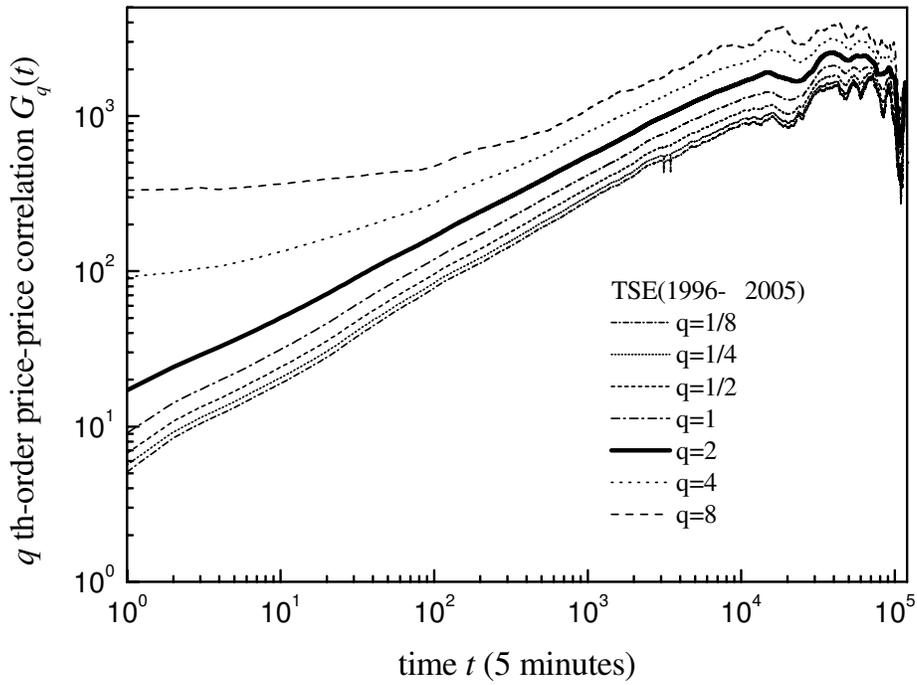}
\caption{\label{fig:3b} Log-log plot of the generalized
price-price correlation function $G_q(t)$ vs. $t$
corresponding to 5 minutes data of the daily Taiwan
stock price index (TSE) from 1996 to 2005. }
\end{figure}

\begin{figure}
\includegraphics{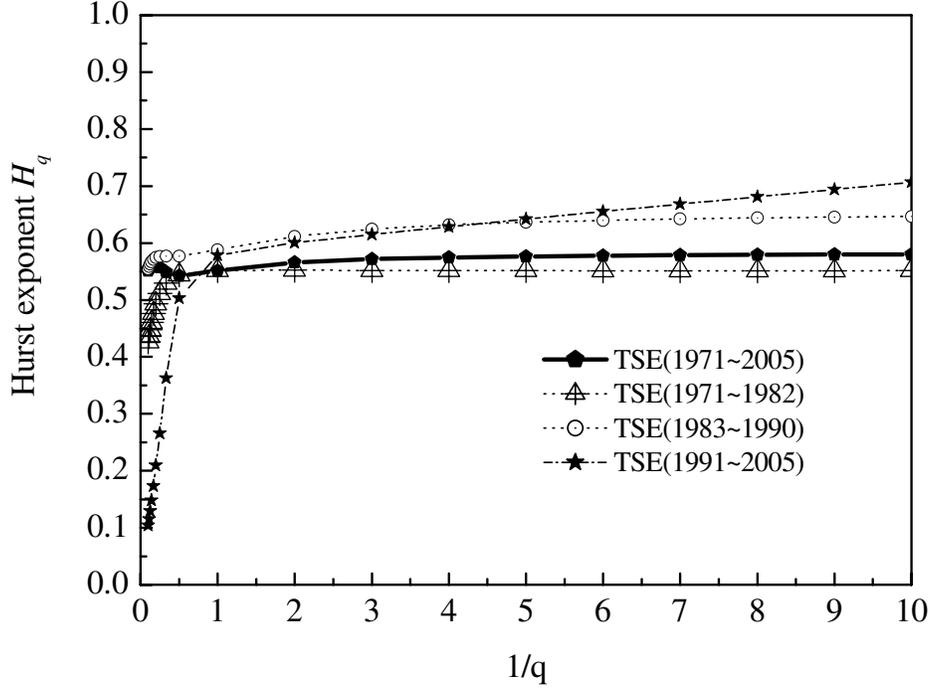}
\caption{\label{fig:4a} Plot of the generalized Hurst
exponent $H_q$ vs. $1/q$ on differ daily Taiwan stock exchange indices.}
\end{figure}

The generalized fluctuation function is defined as
\begin{eqnarray}
F_s^2(\nu)=[{1\over s}\sum_{j=1}^s X_{s}]^2 .
\label{7}
\end{eqnarray}
The generalized $q$th-order fluctuation function is defined as
\begin{eqnarray}
F_q(s)=[{1\over 2N}\sum_{\nu=1}^{2N_s} F_s^2(\nu)^{q/2}]^{1/q}.
\label{8}
\end{eqnarray}

\begin{figure}
\includegraphics{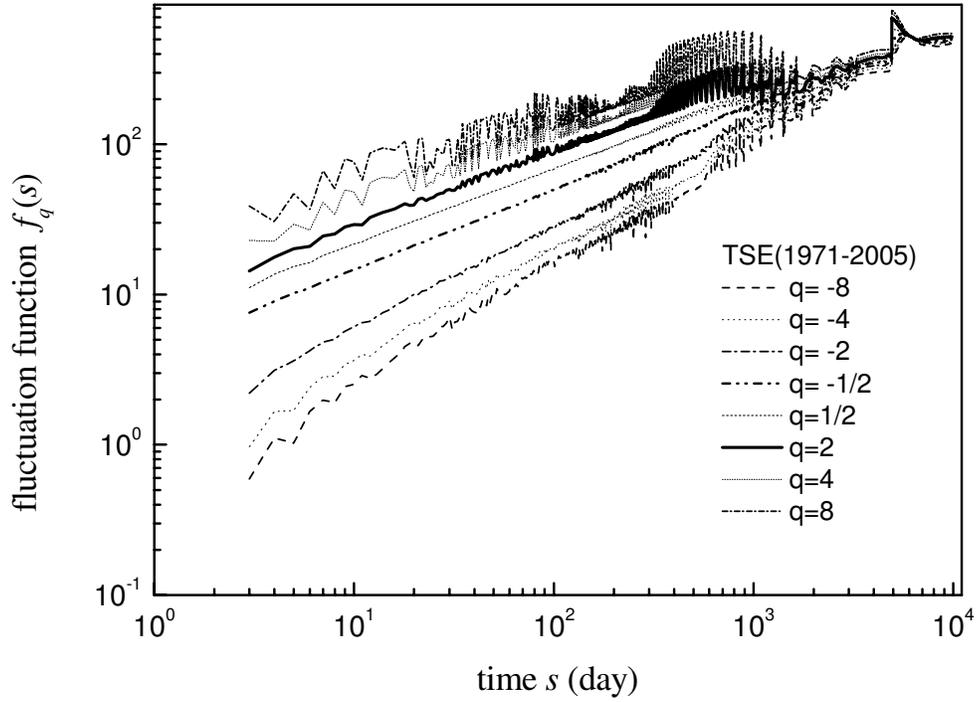}
\caption{\label{fig:5a} Log-log plot of the normalized
fluctuation function $f_q(s)$ vs. $s$ corresponding to the
daily Taiwan stock exchange index (TSE) from 1971 to 2005. }
\end{figure}

\begin{figure}
\includegraphics{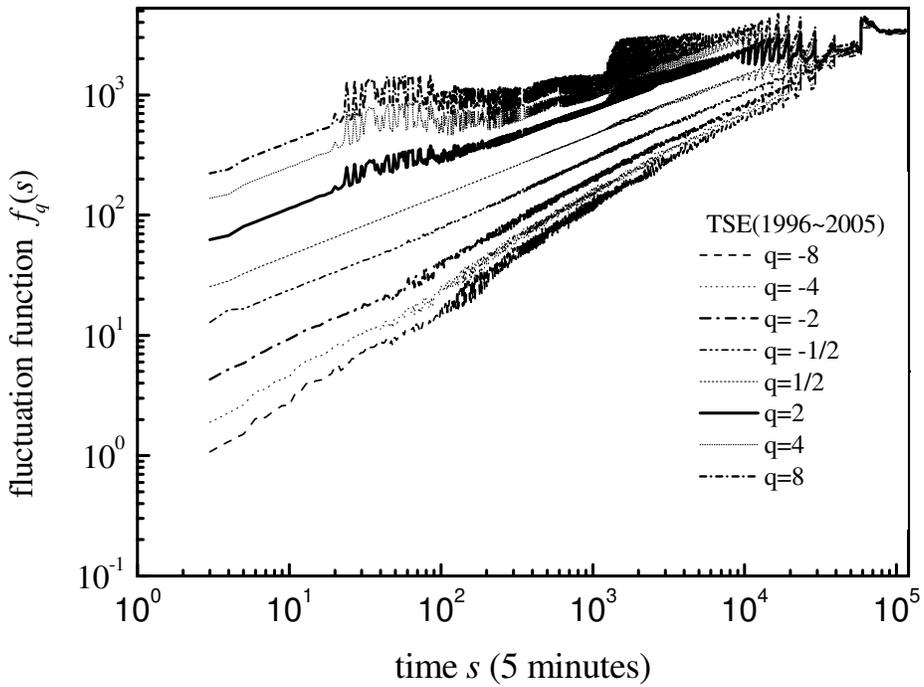}
\caption{\label{fig:5b} Log-log plot of the normalized
fluctuation function $f_q(s)$ vs. $s$ corresponding to 5
minutes data of the daily Taiwan stock price index (TSE)
from 1996 to 2005.}
\end{figure}

\begin{figure}
\includegraphics{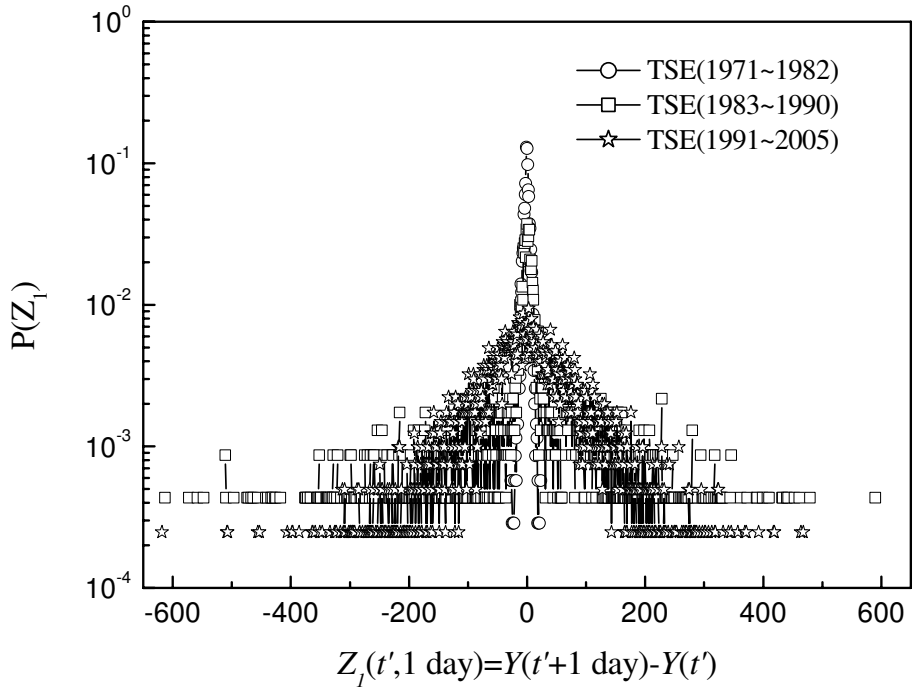}
\caption{\label{fig:6b} Plot of the $P(Z_1)$ vs. $Z_1(t')$ differ
time data on the daily Taiwan stock exchange index (TSE).}
\end{figure}

By construction, since we use a linear fit for simplicity,
$F_q(s)$ is defined for $s\geq 3$. The scaling form of the correlation
function $F_q(s) \sim H_q$ provides the family of generalized
Hurst exponents $H_q$. For reasons that will become clearer
very shortly we also introduce the dimensionless fluctuation
function is defined as
\begin{eqnarray}
f_q(s)={[F_s^2(\nu)^{q/2}]^{1/2}\over [{1\over N}\sum_k^i (r_{k}-{\bar r})^2]^{1/2}}.
\label{9}
\end{eqnarray}
Let us denote the value of the TSE at a certain
time $t'$ as $Y(t')$. We calculate the pdf $P(Z_1)$ of the index changes
\begin{eqnarray}
Z_1(t')=Y(t'+\Delta t)-Y(t').
\label{10}
\end{eqnarray}

\section{\label{sec:level3}Discuss and Results}

A "persistence exponent" $\theta_p$ is defined for
non-equilibrium critical phenomena. Based on large
amounts of data compiled in past years, especially
those records in minutes or seconds, it becomes
possible to perform relatively accurate analysis
and to study the fine structure of the dynamics. In
Fig.~\ref{fig:2a} and \ref{fig:2b},
we observe the persistence probability that at least up to
least up to 250. The slope of persistence probability $P(t)$
estimation from the initial time to 200.

The price evolution is multifractal if the exponent
hierarchy $H_q$ varies with $q$, otherwise is fractal in
the theory of surface dynamical scaling referred to
as multiaffine and self-affine, respectively. In particular,
for $q = 2$, we recover the fractional Brownian motion
case described by the well-known Hurst exponent, $0 < H_2 < 1$.
The bridge between these two analyses is provided by
the second-order Hurst exponent $H_2$ associated with the
correlation function of the stock price, which has been
shown to be simply related to the persistence exponent
through $H_2 = 1-\theta_p$.

We note that this relation holds for any zero-mean
process (not necessarily Gaussian \cite{Hansen,Maslov})that satisfies
requirements above.

\begin{table}
\caption{\label{tab:table1}Compare $H_2$ , $\theta_p$ value
on the daily Taiwan stock exchange index (TSE).  }
\begin{ruledtabular}
\begin{tabular}{lcr}
Time & $H_2$ & $\theta_p$\\
\hline
1971 - 1982 (daily) & 0.54 & 0.41\\
1983 - 1990 (daily) & 0.58 & 0.34\\
1991 - 2005 (daily) & 0.50 & 0.42\\
1971 - 2005 (daily) & 0.54 & 0.46\\
1996 - 2005 (5 minutes) & 0.52 & 0.46\\
\end{tabular}
\end{ruledtabular}
\end{table}

\section{\label{sec:level4}Conclusion}

We analyze the daily Taiwan stock exchange index
(TSE) from 1971 to 2005 and the 5-min intraday data
from 1996 to 2005. The persistence exponent $\theta_p$
associated with the power-law decay of the average
probability.

Our studies base on the persistence probability
analysis of the critical behavior in an economic
index, and the numerical estimation of the persistence
exponent $\theta_p$  with Hurst exponent $H_2$.


\end{document}